\documentclass{moriond}
\usepackage{amssymb,amsmath}

\graphicspath{{figures/}}

\def\Journal#1#2#3#4{{#1} {\bf #2}, #3 (#4)}

\def\PLB{{\em Phys. Lett.}  B}
\def\PRL{\em Phys. Rev. Lett.}
\def\PRD{{\em Phys. Rev.} D}

\def\be{\begin{equation}}
\def\ee{\end{equation}}
\def\bea{\begin{eqnarray}}
\def\eea{\end{eqnarray}}

\newcommand{\Photo}{\includegraphics[height=35mm]{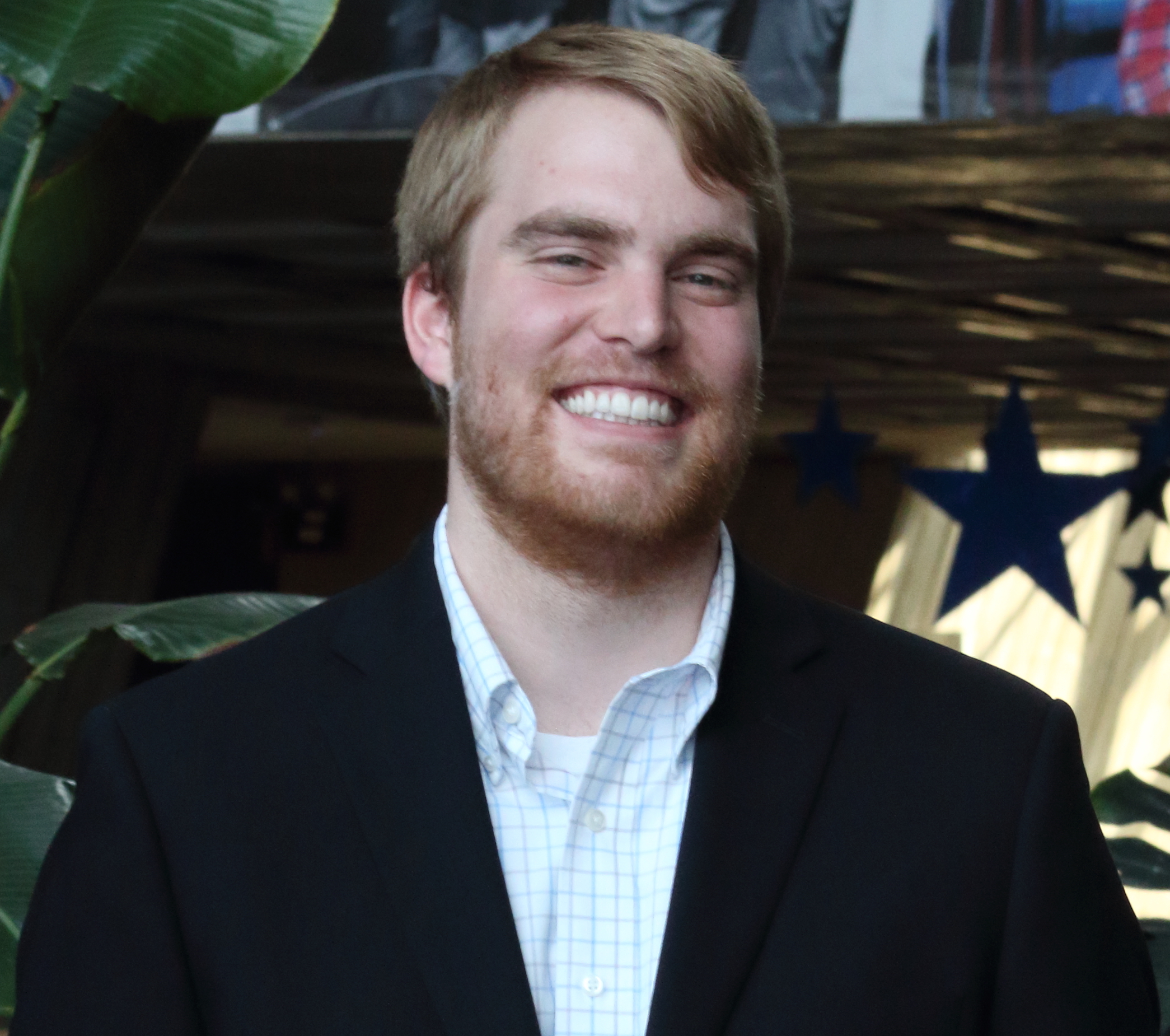}}

\begin{document}
\vspace*{4cm}
\title{NEW CONSTRAINTS ON STERILE NEUTRINOS WITH MINOS/MINOS+ AND DAYA BAY}

\author{ THOMAS JOSEPH CARROLL }

\address{Department of Physics, The University of Texas at Austin, 2515 Speedway, C1600\\
Austin, Texas 78712-1192, USA}

\maketitle\abstracts{
I describe two new searches for sterile neutrino oscillations from the MINOS and Daya Bay experiments. MINOS looked for evidence through muon neutrino disappearance with data collected from the NuMI neutrino beam. Daya Bay searched for evidence through electron antineutrino disappearance using data collected from nuclear reactors. I explain how the MINOS and Daya Bay searches were combined to produce constraints on the same phase space as LSND and MiniBooNE. Finally, I present the status of the sterile neutrino search using data from MINOS+.}

\section{Introduction}

The mixing of three neutrino states is experimentally well established.\,\cite{PDG} This mixing is described by the $3 \times 3$ Pontecorvo-Maki-Nakagawa-Sakata (PMNS) matrix\,\cite{MNS,Pontecorvo1,Pontecorvo2} which can be parameterized~\cite{mixingangles} by three mixing angles $\theta_{12}$, $\theta_{23}$, $\theta_{13}$, and a CP violating phase $\delta$. The oscillation probabilities can be expressed such that they additionally depend on two mass-squared differences $\Delta m^2_{21}$ and $\Delta m^2_{32}$ where $\Delta m^2_{ij} = m^2_i - m^2_j$. However, there are several anomalies\,\cite{reactoranomaly,sageGallex1,sageGallex2,LSND,MiniBooNE} that suggest a mass-splitting inconsistent with those measured assuming the three-flavor paradigm. In particular, the Liquid Scintillator Neutrino Detector (LSND)\,\cite{LSND} and MiniBooNE\,\cite{MiniBooNE} short-baseline neutrino oscillation experiments observed an excess of $\overline{\nu}_e$ events from a $\overline{\nu}_\mu$ beam. Furthermore, results from LEP are consistent with only three light active neutrinos coupled to the Z$^0$ boson based on its invisible decay width.\,\cite{lepz} Thus, one way to address these anomalies is to use a model with three active neutrinos plus one sterile neutrino that does not interact via the weak force. This ``$3+1$" model extends the PMNS matrix by adding one new flavor eigenstate and one new mass eigenstate. The mixing terms can then be parameterized\,\cite{mixingangles} such that, in addition to the original three-flavor parameters, there are three new mixing angles $\theta_{14}$, $\theta_{24}$, $\theta_{34}$ and two new CP violating phases  $\delta_{14}$ and $\delta_{24}$ with $\delta \equiv \delta_{13}$. The oscillation probabilities then require one new mass-squared difference, commonly $\Delta m^2_{41}$.

\section{The MINOS Experiment}

MINOS was an on-axis long-baseline neutrino oscillation experiment that was exposed to the the NuMI neutrino beam from Fermilab. It used a near detector (ND) with a mass of 0.98~kt located 1.04~km from the NuMI target and a far detector (FD) with a mass of 5.4~kt located 735~km from the target. These detectors were functionally equivalent magnetized steel-scintillator, tracking-sampling calorimeters. The detectors consisted of alternating planes of 2.54~cm thick steel plates and 1~cm thick polystyrene-based scintillator strips. Each detector was magnetized by a coil that ran parallel to the length of the detector. The magnetic field allowed the MINOS detectors to distinguish between $\nu_\mu$ and $\overline{\nu}_\mu$ charged-current (CC) interactions based on the curvature of the resulting muon.\cite{MINOSdetectors}

The NuMI beam is produced by colliding 120~GeV protons into a graphite target. The resulting pions and kaons are then focused by two magnetic horns into a decay pipe. The magnetic horns allow the beam to be operated in either a $\nu_\mu$ or $\overline{\nu}_\mu$ mode.
MINOS and MINOS+ collected 11 years of beam data from 2005 to 2016 using the MINOS detectors. The neutrino flux peaked at 3~GeV for MINOS and 7~GeV for MINOS+. In June 2016 the NuMI beam achieved a beam power of 700~kW making it the most powerful neutrino beamline.\cite{beampaper}

\section{The MINOS 3\,+\,1 Sterile Neutrino Analysis}

MINOS has made precision measurements of the three-flavor atmospheric oscillation parameters $\Delta m^2_{32}$ and $\theta_{23}$.\,\cite{minosstdosc} For the $3+1$ model, MINOS is sensitive to $\Delta m^2_{41}$, $\theta_{24}$, and $\theta_{34}$ through muon neutrino disappearance. This analysis studied muon neutrino disappearance using CC and neutral-current (NC) interactions. The sensitivity of MINOS can be illustrated by considering the leading order approximations for the probabilities associated with the analysis channels in this model. The $\nu_\mu$ survival probability is measured with CC interactions and can be written:
\be \label{eq:numu survival}
P(\nu_\mu\rightarrow\nu_\mu) \approx 1 - \sin^22\theta_{23}\cos^2\theta_{24}\sin^2\Delta_{31} - \sin^22\theta_{24}\sin^2\Delta_{41},
\ee
where $\Delta_{ij} = (\Delta m^2_{ij}L/4E)$, $L$ is the distance traveled by the neutrino, and $E$ is the neutrino energy. Equation \ref{eq:numu survival} shows that the CC channel is sensitive to $\theta_{24}$. The addition of a sterile neutrino allows there to be disappearance of NC events expressed as:
\be \label{eq:nc survival}
1  - P(\nu_\mu \rightarrow \nu_s) \approx 1 - c^4_{14}c^2_{34}\sin^22\theta_{24}\sin^2\Delta_{41} - A \sin^2\Delta_{31} + B \sin2\Delta_{31},
\ee
where $c_{ij} = \cos\theta_{ij}$ and $s_{ij} = \sin\theta_{ij}$. The terms $A$ and $B$ are functions of the mixing angles and phases. To first order, $A = s^2_{34}\sin^22\theta_{23}$ and $B = \frac{1}{2}\sin\delta_{24}s_{24}\sin2\theta_{34}\sin2\theta_{23}$. From Eq.~\ref{eq:nc survival}, the NC channel is dependent on the parameters $\theta_{24}$, $\theta_{34}$, and $\delta_{24}$. However, the sensitivity is limited by poor neutrino-energy resolution due to the undetected outgoing neutrino, a lower event rate due to cross sections, and $\nu_\mu$ and $\nu_e$ CC backgrounds. Although $\theta_{14}$ appears in Eq.~\ref{eq:nc survival}, an analysis of solar and reactor neutrino data yields the constraint $\sin^2\theta_{14} < 0.041$ at 90\%~C.L.\,\cite{minosprl33} which is small enough to set $\theta_{14} = 0$ in this analysis. 

\subsection{Event Selection}
The MINOS sterile analysis required the selection of samples of NC and CC $\nu_\mu$ events. This analysis selected events from a beam exposure of $10.56 \times 10^{20}$ protons on target (POT).

NC events have no flavor information and are characterized by a hadronic shower in the detector. These events were selected based on event topology by searching for interactions that induced activity spread over less than 47 steel-scintillator planes. If events had a reconstructed track, then the track was required to penetrate no more than five detector planes beyond the end of the hadronic shower. The NC selection had an efficiency of 79.9\% for the ND and resulted in a sample with a purity of 58.9\%, both estimated from Monte Carlo (MC) simulation. For the FD, assuming standard three-flavor oscillations, the efficiency of the selection was 87.6\% and the sample purity was 61.3\%.

CC $\nu_\mu$ events are characterized by a long muon track that is bending in the magnetic field of the detector and a hadronic shower near the interaction point. A $k$-nearest neighbor algorithm was developed to select these events based on muon track features resulting in a high purity sample.\,\cite{RustemThesis} The algorithm used four variables: the number of detector planes hit by the muon track, the average energy deposited per scintillator plane by the track, the track's transverse energy deposition profile, and the variation of the energy deposited along the muon track. Events were required to have failed the NC selection procedure to be included in the CC sample. The CC selection had an efficiency of 53.9\% for the ND and produced a sample with a purity of 98.7\%, both estimated from MC simulation. For the FD, assuming three-flavor oscillations, the corresponding efficiency was 84.6\% and the purity was 99.1\%.

\subsection{Analysis Technique for the 3\,+\,1 Sterile Neutrino Model}
\begin{figure}[h!]

\centerline{\includegraphics{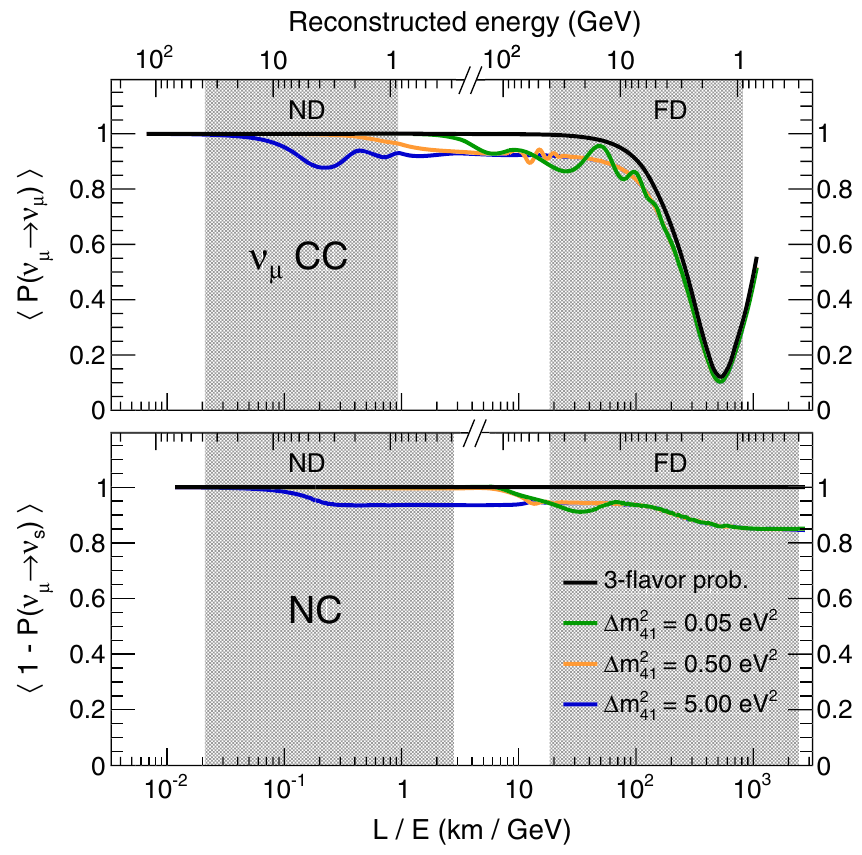}}

\caption[]{Muon neutrino oscillation probabilities as a a function of $L/E$, where $L$ is the distance traveled by the neutrinos, and $E$ is the reconstructed neutrino energy, for three different values of $\Delta m^2_{41}$, with $\theta_{14} = 0.15$, $\theta_{24} = 0.2$, $\theta_{34} = 0.5$, and the values of $\Delta m^2_{31}$, $\Delta m^2_{21}$, $\theta_{12}$, $\theta_{23}$, and $\theta_{13}$ from the Particle Data Group\,\cite{PDG}. The dip in $P(\nu_\mu \rightarrow \nu_\mu)$ at 500~km/GeV is due to oscillations driven by $\Delta m^2_{31}$. As $L/E$ increases, the various oscillation probabilities become similar and the lines overlap. The gray bands indicate the regions of reconstructed energy where CC $\nu_\mu$ interactions (top panel) and NC interactions (bottom panel) are observed in the two MINOS detectors.}
\label{fig:osc prob L/E}
\end{figure}

The MINOS analysis used CC $\nu_\mu$ and NC events to look for perturbations on three-flavor oscillations. Figure \ref{fig:osc prob L/E} shows examples for different values of $\Delta m^2_{41}$ and how they alter the oscillation probabilities in both channels at the MINOS detectors. For $10^{-3} \lesssim \Delta m^2_{41} \lesssim 0.1$~eV$^2$ an energy-dependent depletion of $\nu_\mu$ events would be observed only at the FD. For $0.1 \lesssim \Delta m^2_{41} \lesssim 1$~eV$^2$ fast oscillations occur at the FD that are averaged out due to the energy resolution of the detector leading to a constant deficit of events. For $1 \lesssim \Delta m^2_{41} \lesssim 100$~eV$^2$ an energy-dependent depletion of $\nu_\mu$ events would be seen at the ND with fast oscillations being averaged out at the FD. Then for $\Delta m^2_{41} \gtrsim 100$~eV$^2$ oscillations occur upstream of the ND leading to event deficits in both detectors. The possibility for oscillations at the ND\,\cite{HernandezYu} means that the ND spectrum cannot be used to predict the FD spectrum as was traditionally done in MINOS oscillation analyses\,\cite{MINOSprd}.
In order to be sensitive to oscillations at the ND, MINOS analyzed the ratio of the FD energy spectrum to the ND energy spectrum. The oscillated Far-over-Near MC energy spectrum ratio is then fit to the Far-over-Near data energy spectrum ratio. 

\subsection{Fitting Procedure}
\begin{figure}[h!]
\centerline{\includegraphics[height=0.55\textwidth]{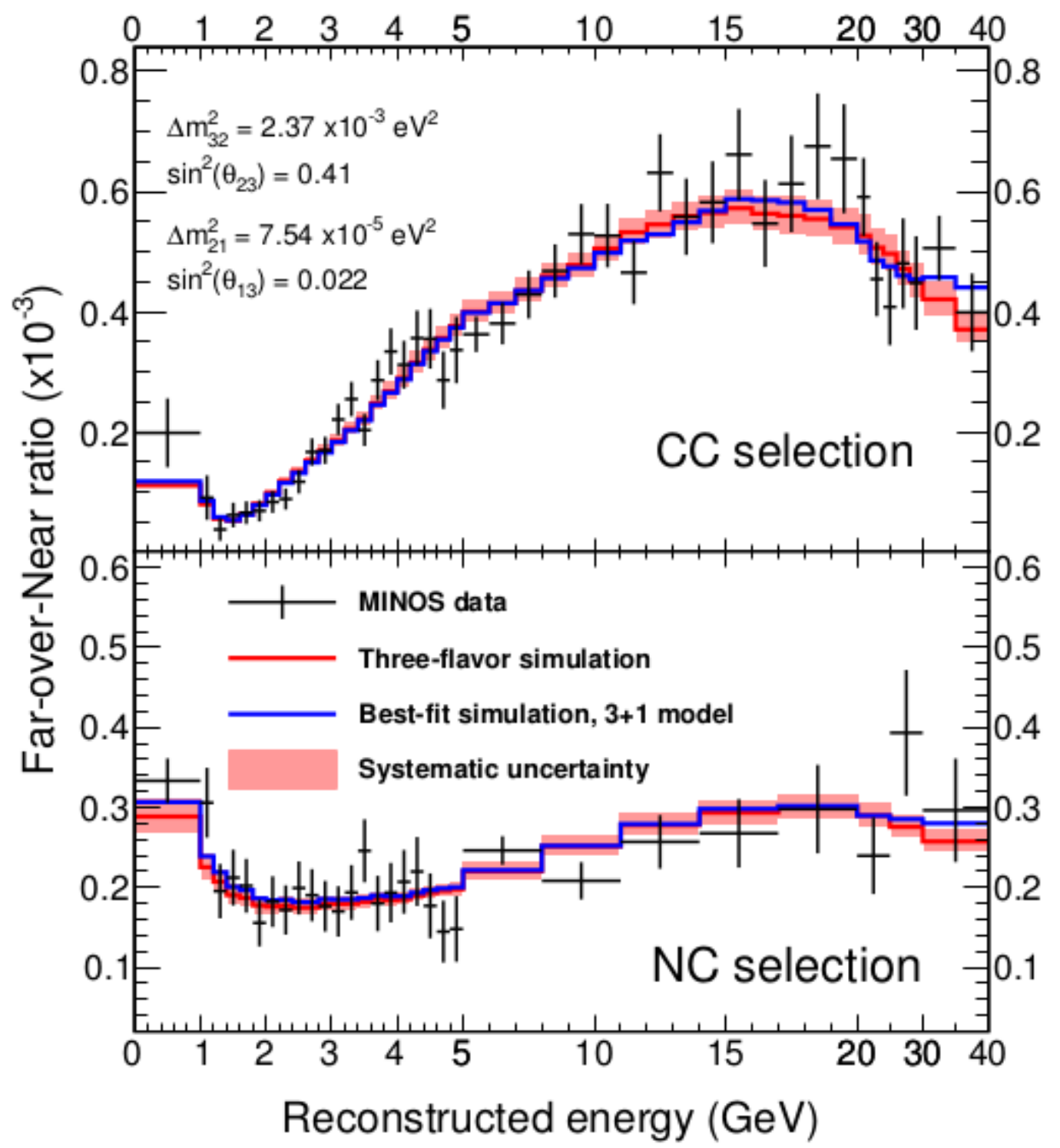}}
\caption[]{The ratios of the energy spectra in the MINOS FD to those in the ND, shown for the CC $\nu_\mu$ (top) and NC (bottom) samples. The solid lines represent the predicted ratios from fits to the standard three-flavor oscillation model (red) and to the $3+1$ sterile neutrino model (blue).}
\label{fig:CC and NC ratios}
\end{figure}

The CC and NC spectra ratios were fit simultaneously using the exact oscillation probabilities to determine $\theta_{23}$, $\theta_{24}$, $\theta_{34}$, $\Delta m^2_{32}$, and $\Delta m^2_{41}$. MINOS is not sensitive to $\delta_{13}$, $\delta_{14}$, $\delta_{24}$, and $\theta_{14}$. Therefore, all were set to zero. The values $\sin^2\theta_{12} = 0.307$ and $\Delta m^2_{21} = 7.54 \times 10^{-5}$~eV$^2$ were set based on a global fit to neutrino data\,\cite{solarfit}, and $\sin^2\theta_{13} = 0.022$ based on a weighted average of recent results from reactor experiments\,\cite{dayabayt13,renot13,dchoozt13}. Figure \ref{fig:CC and NC ratios} shows good agreement between the Far-over-Near ratios measured and predicted using a three-flavor hypothesis.

The fit minimized the $\chi^2$ function in Eq.~\ref{eq:minoschi2} where $x_m$ is the observed ratio in bin $m$, $\mu_m$ is the predicted ratio, and $V$ is an $N \times N$ covariance matrix expressing the statistical and systematic uncertainties of the predicted ratio.  The second term in Eq.~\ref{eq:minoschi2} is a flux penalty term where $X$ and $M$ are the observed and predicted total number of events in the ND, and $\sigma_{M}$ is conservatively set to 50\% of $M$ based on measurements of the NuMI beam muon flux.
\be \label{eq:minoschi2}
\chi^2_{\rm CC, NC} = \sum_{m=1}^N\sum_{n=1}^N(x_m - \mu_m)(V^{-1})_{mn}(x_n - \mu_n) + \frac{(X - M)^2}{\sigma^2_M}
\ee

\subsection{Systematic Uncertainties}
\begin{figure}[h]

\centerline{\includegraphics[height=0.5\textwidth]{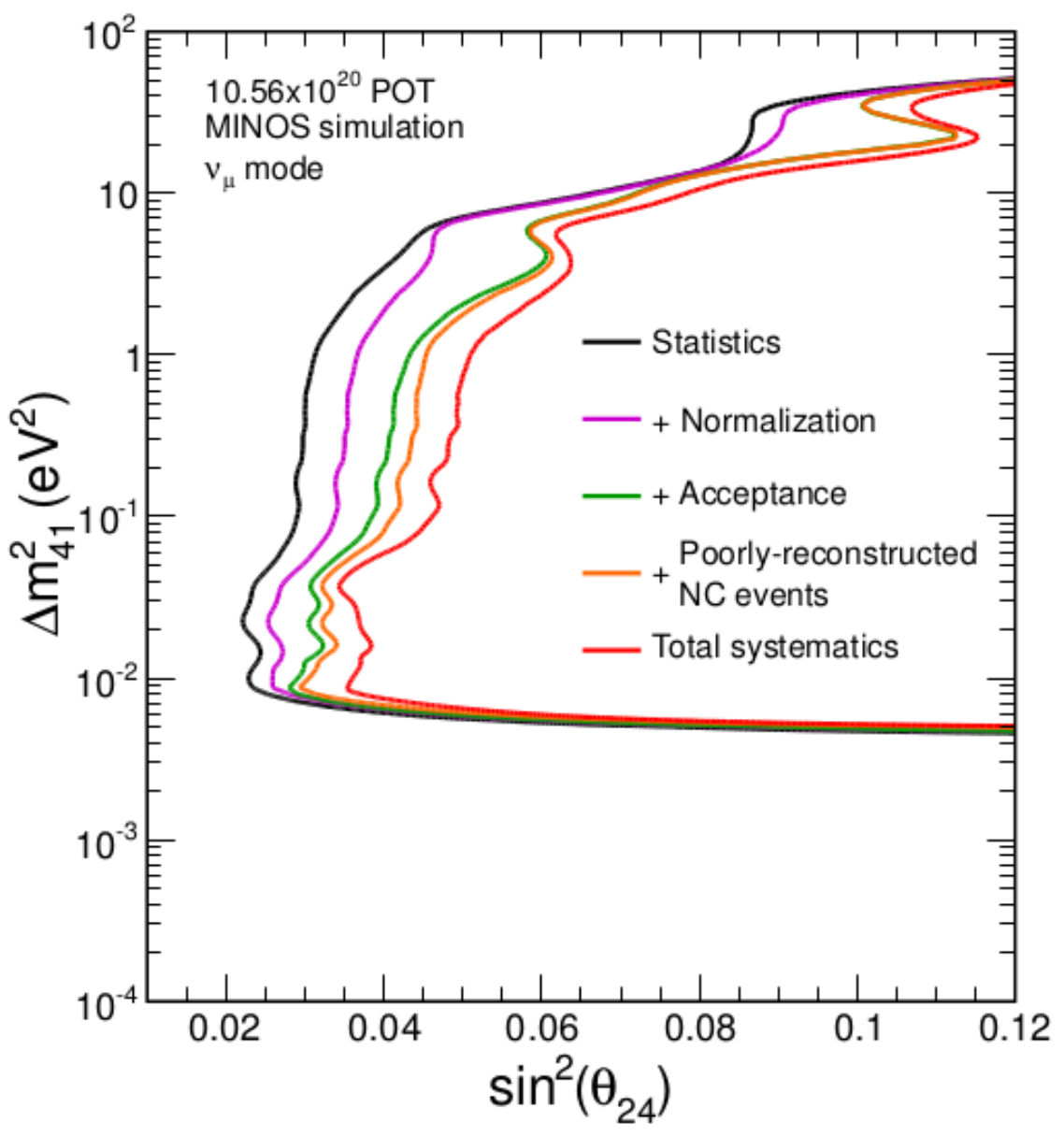}}
\caption[]{The effects of systematic uncertainties on the MINOS 90\% C.L. sensitivity in the $(\sin^2\theta_{24}, \Delta m^2_{41})$ plane, shown by successive inclusion of the listed uncertainties.}
\label{fig:MINOS systematics}
\end{figure}

The covariance matrix in Eq.~\ref{eq:minoschi2} can be broken down into its component uncertainties as:
\be \label{eq:covar}
V = V_{\rm stat} + V_{\rm norm} + V_{\rm acc} + V_{\rm NC} + V_{\rm other}.
\ee
Figure \ref{fig:MINOS systematics} shows the effect of incrementally adding the systematic uncertainties to the sensitivity. $V_{\rm stat}$ contains the statistical uncertainty. $V_{\rm norm}$ contains the uncertainty in the relative normalization of the CC and NC samples between the ND and FD which accounts for uncertainties in reconstruction efficiencies. $V_{\rm acc}$ accounts for uncertainties on the acceptance and selection efficiency of the ND. This systematic uncertainty has the largest effect on the sensitivity as seen in Fig.~\ref{fig:MINOS systematics} due to the fact that it is only for the ND and thus cannot be canceled out by the FD. These uncertainties were evaluated by varying event selection requirements in the data and MC simulation to probe known weaknesses in the simulation. As these requirements were varied, the total variations in the ND data to MC ratios were taken as systematic uncertainties on the Far-over-Near ratios. $V_{\rm NC}$ accounts for the uncertainty on the procedure used to remove poorly reconstructed events from the NC sample. $V_{\rm other}$ includes terms to account for all sources of uncertainty in neutrino interaction cross sections and the flux of neutrinos produced in the NuMI beam.\,\cite{MINOSsterilePRL}

\subsection{MINOS 3\,+\,1 Model Limit}
\begin{figure}[h]

\centerline{\includegraphics[height=0.5\textwidth]{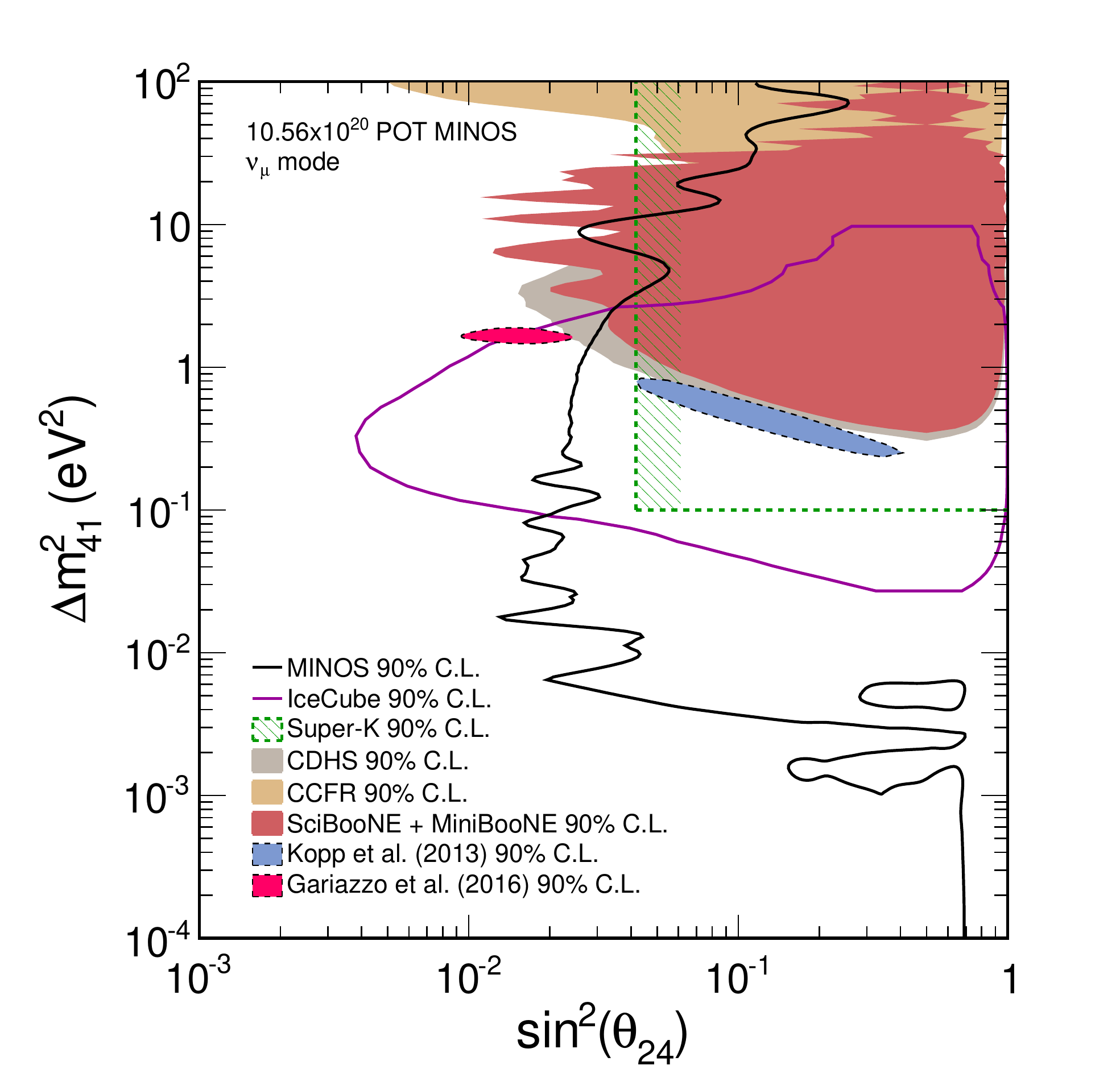}}
\caption[]{The MINOS 90\% confidence limit in the $(\sin^2\theta_{24}, \Delta m^2_{41})$ plane compared to results from other experiments\,\cite{icecubelimit,superklimit,CDHSlimit,CCFRlimit,boonelimit} and global fits\,\cite{Kopp,Gariazzo}. To compare these global fits to disappearance data, $\sin^{2}2\theta_{\mu e}$ is converted to $\sin^{2}\theta_{24}$ by assuming $\theta_{14}$ = 0.15, the best fit value a global fit to appearance data\,\cite{Kopp}.}
\label{fig:MINOS 90CL}
\end{figure}

Since the MINOS best fit was consistent with three-flavor oscillations, the data can be used to set a muon neutrino disappearance limit. The limit was set by dividing the $(\sin^2\theta_{24}, \Delta m^2_{41})$ plane into fine bins and minimizing Eq.~\ref{eq:minoschi2} at each bin allowing $\Delta m^2_{32}$, $\theta_{23}$, and $\theta_{34}$ to vary. The significance of the $\Delta\chi^2$ with respect to the global minimum was calculated using the Feldman-Cousins method\,\cite{FCmethod}. The resulting MINOS 90\% C.L.\,\cite{MINOSsterilePRL} is shown in Fig.~\ref{fig:MINOS 90CL}. It excludes a sterile neutrino over six orders of magnitude in $\Delta m^2_{41}$ and two orders of magnitude in $\sin^2\theta_{24}$. The MINOS limit is the best constraint below 0.1~eV$^2$ in this phase space.
Below $\Delta m^2_{41} = 10^{-2}$~eV$^2$ there is an internal allowed region and a feature near $\Delta m^2_{41} = 2 \times 10^{-3}$~eV$^2$ which are due to expected degenerate predictions with the three-flavor case.

\section{MINOS, Daya Bay Combination}

MINOS measured muon neutrino disappearance and thus can measure the matrix element $|U_{\mu4}|^2$. However, the LSND and MiniBooNE experiments measured muon neutrino to electron neutrino appearance and constrainted allowed values of $\sin^22\theta_{\mu e}$ which is defined by the matrix elements $|U_{e4}|^2$ and $|U_{\mu4}|^2$. Assuming CPT conservation,  a muon neutrino disappearance measurement must be combined with an electron neutrino disappearance measurement, which can measure the matrix element $|U_{e4}|^2$, in order to look at the same mixing angle as LSND and MiniBooNE.

\subsection{Daya Bay}
To constrain the same phase space as LSND and MiniBooNE, the MINOS measurement was combined with the Daya Bay reactor neutrino experiment\,\cite{DayaBayNIM}. Daya Bay uses eight identical detectors to measure intense sources of neutrinos from six reactor cores with a total power of 17.4~GW$_{\rm th}$. The detectors are arranged with two at both near experimental halls and four at the far experimental hall. Daya Bay detects electron antineutrinos via inverse beta decay (IBD). The main volume of the Daya Bay detectors is linear alkylbenzene-based liquid scintillator doped with gadolinium which increases neutron capture. Daya Bay was designed to measure $\theta_{13}$ and is responsible for the most precise measurement of electron antineutrino disappearance to date\,\cite{DayaBayPRD}.

\subsection{Daya Bay Sterile Neutrino Fit and Limit}

\begin{figure}[h!]

\centerline{\includegraphics[height=0.5\textwidth]{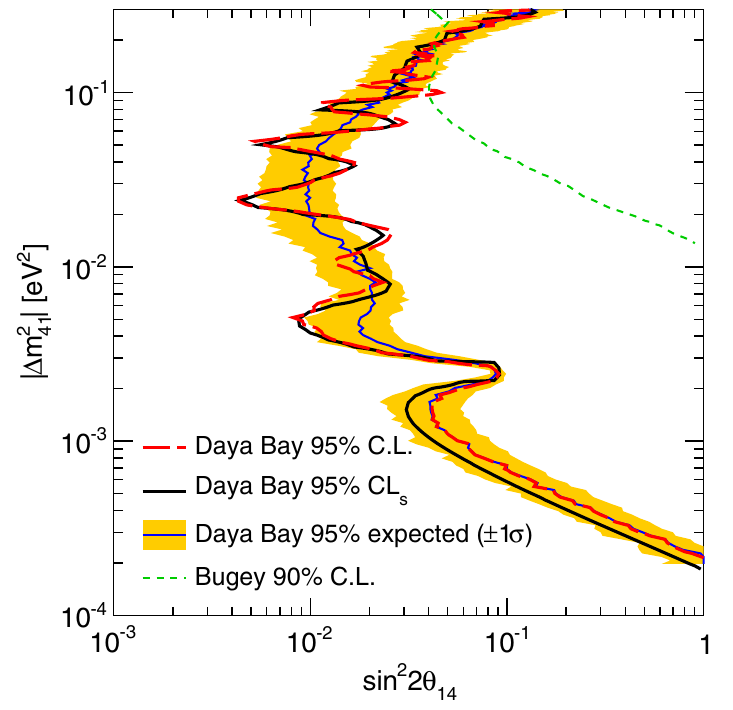}}
\caption[]{The Daya Bay limits in the $(\sin^2\theta_{14}, |\Delta m^2_{41}|)$ plane under the assumption of $\Delta m^2_{32} > 0$ and $\Delta m^2_{41} > 0$. The red long-dashed curve is the 95\% C.L. exclusion contour with the Feldman-Cousins method\,\cite{FCmethod} from method A. The black solid curve is the 95\% CL$_s$ exclusion contour\,\cite{CLs1} from method B. The expected 95\% C.L. 1$\sigma$ band in yellow is centered around the sensitivity curve, shown as a thin blue line. The Daya Bay limit is compared to Bugey's\,\cite{bugey3} 90\% C.L. limit shown as the green dashed curve.}
\label{fig:Daya Bay limit}
\end{figure}

For this sterile neutrino search, Daya Bay analyzed IBD data from 217 days in a partial configuration using six detectors plus 404 days in the full configuration. This analysis used two different methods, referred to as method A and B, to fit the data. Method A used the energy spectra measured at the near halls to predict the far hall energy spectrum. The fit then minimized a $\chi^2$ function. Method B simultaneously fit all of the spectra from the Daya Bay detectors using the predicted reactor flux constrained by the Huber\,\cite{huber} and Muller\,\cite{muller} models. For this method, the systematic uncertainty on the flux was increased from 2\% to 5\% to cover observed discrepancies with the predicted reactor neutrino spectrum.\footnote{Daya Bay recently performed a detailed study of their reactor antineutrino flux and spectrum.\,\cite{DayaBayFlux}} Method B maximized a log-likelihood function complete with nuisance parameters for systematic uncertainties. Both methods used the exact oscillation probabilities to determine $\theta_{13}$, $\theta_{14}$, and $\Delta m^2_{41}$. For method A, the Feldman-Cousins procedure\,\cite{FCmethod} was used to set limits while method B set limits using the CL$_s$ technique\,\cite{CLs1,CLs2,CLs3}. Daya Bay sets the most stringent limits for $\Delta m^2_{41} \lesssim 0.2$~eV$^2$ in $\sin^22\theta_{14}$. Figure \ref{fig:Daya Bay limit} shows the 95\% C.L. from the Feldman-Cousins method\,\cite{FCmethod} and the 95\% CL$_s$ exclusion contour\,\cite{CLs1}.\,\cite{DayaBaySterilePRL} Methods A and B provide consistent results as seen in Fig.~\ref{fig:Daya Bay limit}. The slight difference seen between the two limits for $\Delta m^2_{41} \lesssim 2 \times 10^{-3}$~eV$^2$ is due to limited statistics relevant for this region which effect the techniques differently.

\subsection{Combined Limit}
\begin{figure}[h!]
\centerline{\includegraphics[height=0.55\textwidth]{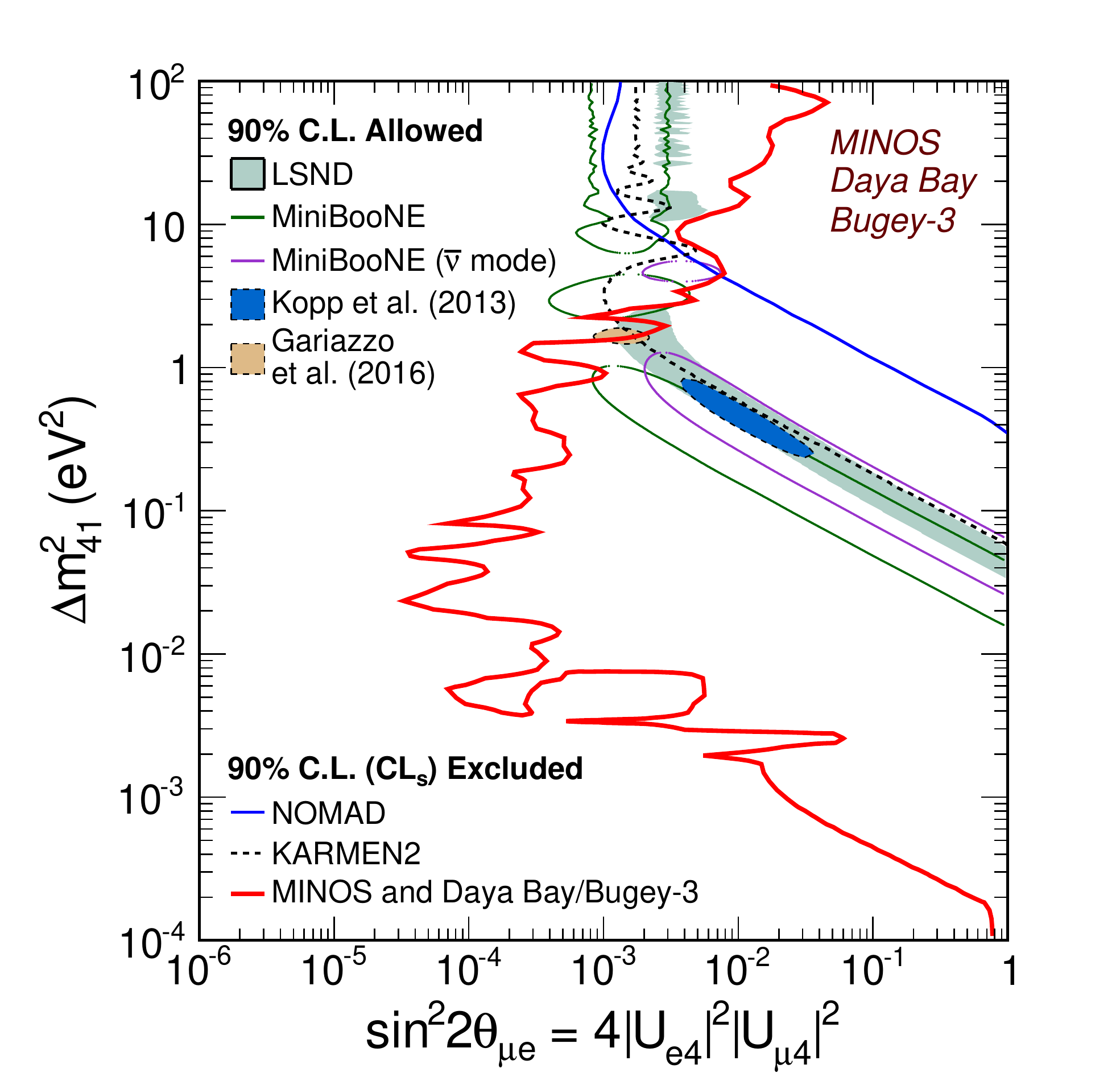}}
\caption[]{MINOS and Daya Bay\,$+$\,Bugey-3 combined 90\% CL$_s$ limit on $\sin^22\theta_{\mu e}$ compared to the LSND\,\cite{LSND} and MiniBooNE\,\cite{MiniBooNE} 90\% C.L. allowed regions. Regions of parameter space to the right of the red contour are excluded. The regions excluded at 90\% C.L. by the KARMEN2 Collaboration\,\cite{karmen} and the NOMAD Collaboration\,\cite{nomad} are also shown. Note that the feature in the exclusion contour near $\Delta m^2_{41} = 5 \times 10^{-3}$~eV$^2$ is due to the island in Fig.~\ref{fig:MINOS 90CL}.}
\label{fig:MINOS + Daya Bay 90CL}
\end{figure}

MINOS and Daya Bay did a combined analysis using a common CL$_{s}$ method\,\cite{CLs1,CLs2}. Before combining with MINOS, the Daya Bay and Bugey-3\,\cite{bugey3} electron antineutrino disappearance measurements were combined taking into account correlated systematic uncertainties. Bugey-3 made measurements at shorter baselines than Daya Bay which provides increased sensitivity for $\Delta m^2_{41} \gtrsim 0.2$~eV$^2$. For the combination of MINOS and Daya Baya + Bugey-3, systematic uncertainties are taken to be uncorrelated. Figure \ref{fig:MINOS + Daya Bay 90CL} shows the combined 90\% CL$_s$ exclusion contour\,\cite{MINOSDayaBayPRL}. The limit constrains $\sin^22\theta_{\mu e}$ over six orders of magnitude in $\Delta m^2_{41}$. This limit is the strongest constraint to date and excludes the sterile neutrino mixing phase space allowed by the LSND and MiniBooNE experiments for $\Delta m^2_{41} < 0.8$~eV$^2$ at a 95\% CL$_s$.

\section{MINOS+}
The increased intensity and beam energy of MINOS+ make it well-suited for sterile neutrino searches. MINOS+ is improving on MINOS with more data and an improved fit technique.

\subsection{First Half of MINOS+ Data}
\begin{figure}[h!]

\centerline{\includegraphics[height=0.5\textwidth]{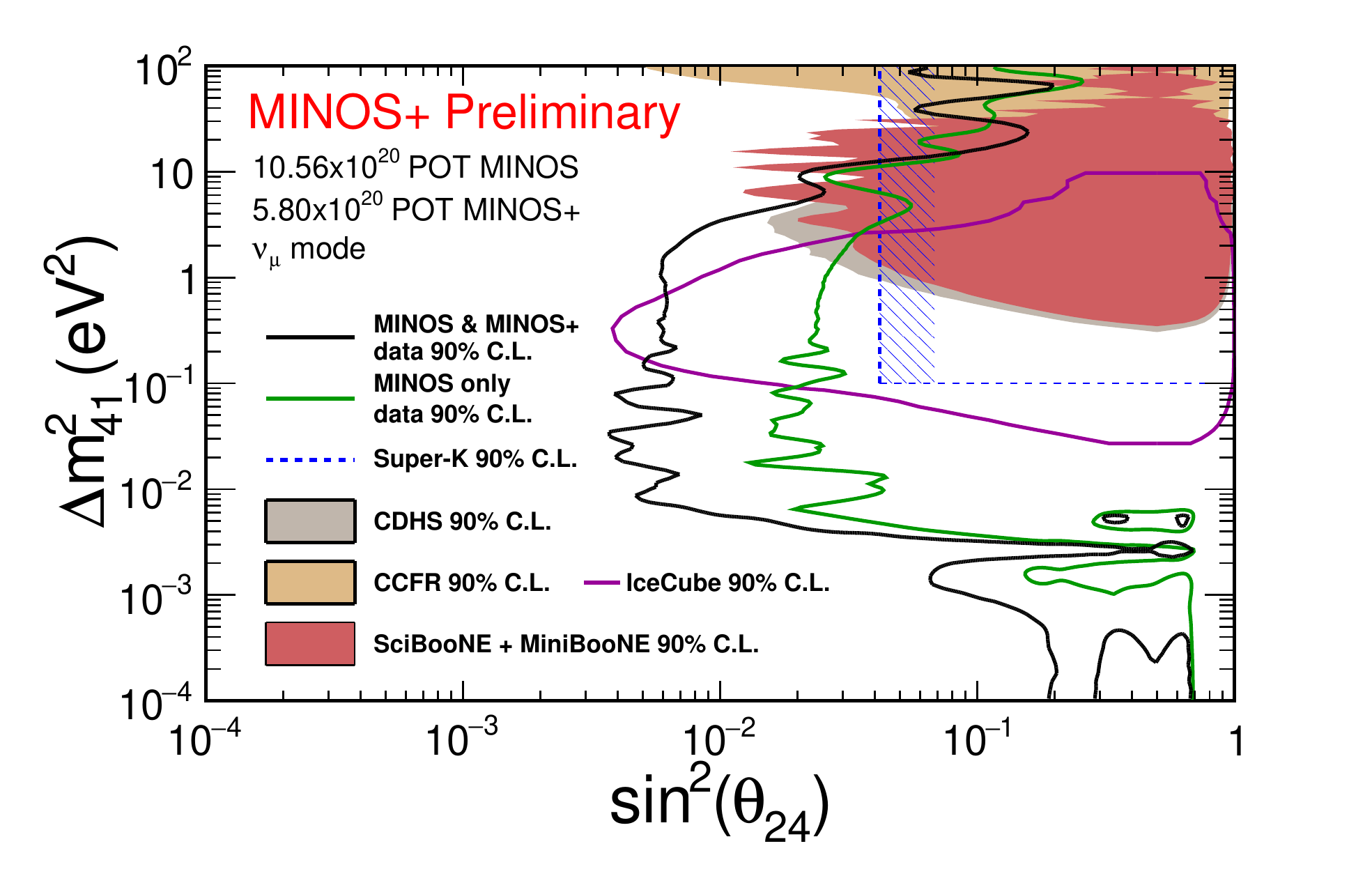}}
\caption[]{The combined MINOS and MINOS+ 90 $\%$ C.L. confidence limit in the $(\sin^2\theta_{24}, \Delta m^2_{41})$ plane compared to the MINOS\,\cite{MINOSsterilePRL} and other experiments\,\cite{icecubelimit,superklimit,CDHSlimit,CCFRlimit,boonelimit}.}
\label{fig:MINOS++}
\end{figure}

The first two years of MINOS+ data represent a beam exposure of $5.80 \times 10^{20}$~POT. When these data are added to the MINOS dataset using the analysis described above there is a significant increase in the exclusion of $\sin^2\theta_{24}$ for $10^{-2} \lesssim \Delta m^2_{41} \lesssim 2$~eV$^2$. This improvement is largely due to the increased beam energy of MINOS+ which provided more statistics at higher neutrino energies compared to MINOS. In Fig.~\ref{fig:MINOS++}, the exclusion limit using MINOS and MINOS+ data is compared to the most recent MINOS limit\,\cite{MINOSsterilePRL}.

\subsection{Two-Detector Fit Technique}
\begin{figure}[h!]

\centerline{\includegraphics[height=0.5\textwidth]{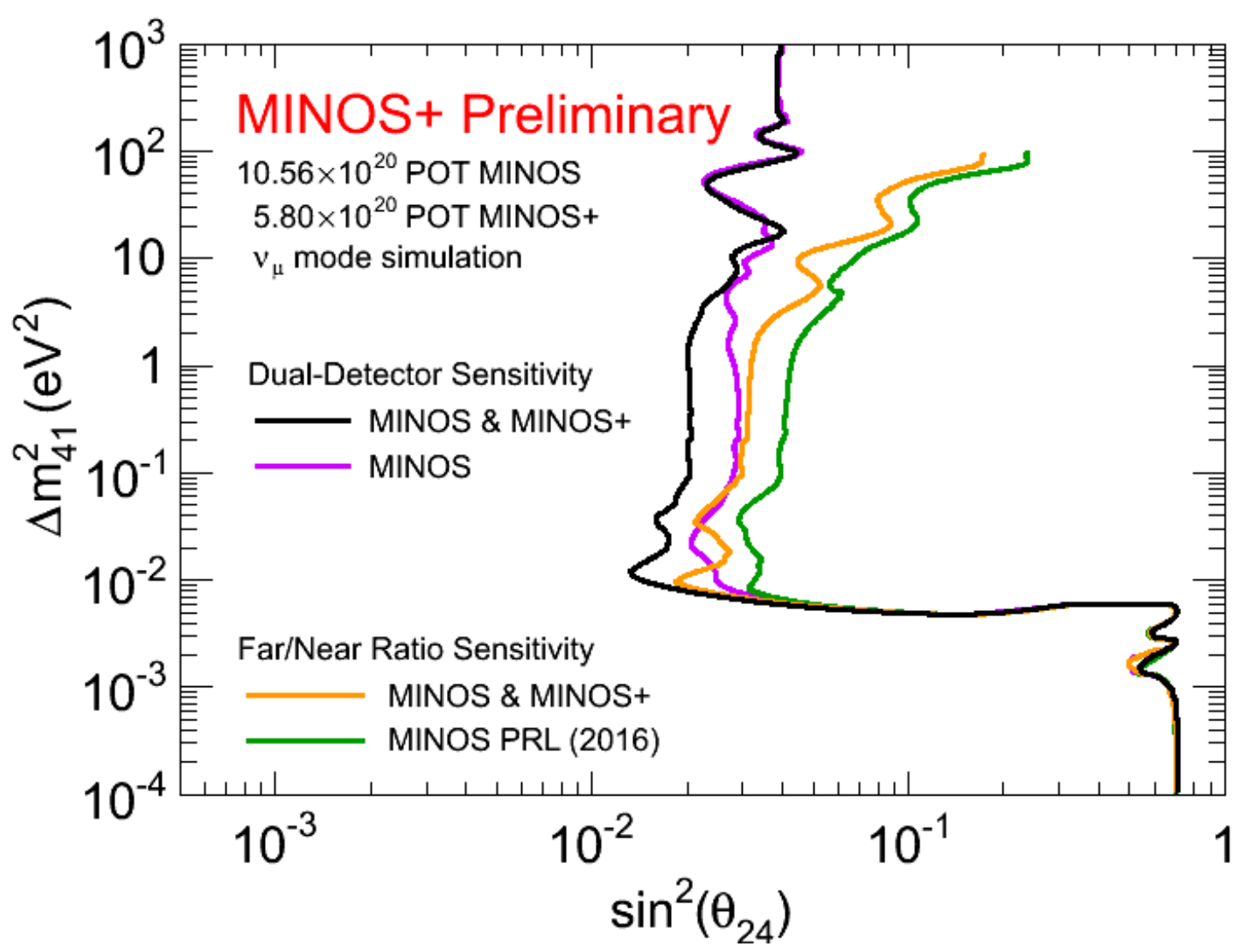}}
\caption[]{The sensitivity at 90\% C.L. for MINOS ($10.56 \times 10^{20}$~ POT) alone and combined with MINOS+ ($5.80 \times 10^{20}$~POT) compared for the two-detector fit method (violet, black) and the Far-over-Near ratio method (green, gold).}
\label{fig:dualfit}
\end{figure}

The MINOS+ analysis is also being improved by fitting the spectra in both detectors simultaneously rather than fitting the ratios. In addition to being being less sensitive to oscillations upstream of the ND, the ratio technique had reduced sensitivity to oscillations at the ND as well as to deficits between the detectors due to the fact that the statistical uncertainty was dominated by the FD. The two-detector fit technique allows the analysis to take full advantage of the large statistics available at the ND and the shape information provided by the spectra from both detectors. For $\Delta m^2_{41} \gtrsim 5$~eV$^2$ the sensitivity is improved by the ND statistics and the ability to compare the ND and FD spectra. These advantages significantly increase the sensitivity of MINOS+ to exclude regions of $\sin^2\theta_{24}$ for $\Delta m^2_{41} > 100$~eV$^2$ as seen in Fig.~\ref{fig:dualfit}. For $10^{-2} \lesssim \Delta m^2_{41} \lesssim 5$~eV$^2$ the sensitivity is improved by the cancellation of uncertainties between the ND and FD. 

\section{Conclusion}
MINOS extended their 90\% C.L. exclusion limit over six orders of magnitude in $\Delta m^2_{41}$. Through close collaboration, Daya Bay and MINOS were able to use the CL$_s$ technique\,\cite{CLs1,CLs2} to combine their disappearance limits to extract equivalent appearance limits, assuming the $3+1$ model. This result increases the tension between appearance and disappearance sterile neutrino searches for $\Delta m^2_{41} < 1$~eV$^2$. These searches will be updated in the future. Daya Bay and MINOS have an agreement for a future combination, and MINOS+ has 50\% more data to analyze.

\section*{Acknowledgments}

This work was supported by the U.S. DOE; the United Kingdom STFC; the U.S. NSF; the State and University of Minnesota; and Brazil's FAPESP, CNPq and CAPES. We are grateful to the Minnesota Department of Natural Resources and the personnel of the Soudan Laboratory and Fermilab. We thank the Texas Advanced Computing Center at The University of Texas at Austin for the provision of computing resources.

\end{document}